\documentclass{article}
\usepackage{graphics}
\usepackage{amsmath}
\usepackage{amssymb}
\usepackage{epsfig}
\usepackage{bm}

\begin{document}

\title{Generalization of multifractal theory \break within quantum calculus}
\maketitle{}
\author{Alexander Olemskoi$^{1,2}$, Irina Shuda$^{2}$, Vadim Borisyuk$^{2}$} \\

\textit{$^{1}$Institute of Applied Physics, Nat. Acad. Sci. of Ukraine, 58,\\
Petropavlovskaya St., 40030 Sumy, Ukraine\\}

\textit{$^{2}$Sumy State University, 2,
Rimskii-Korsakov St., 40007 Sumy, Ukraine}

\begin{abstract} On the basis of the deformed series in quantum calculus, we
generalize the partition function and the mass exponent of a multifractal, as
well as the average of a random variable distributed over self-similar set. For
the partition function, such expansion is shown to be determined by
binomial-type combinations of the Tsallis entropies related to manifold
deformations, while the mass exponent expansion generalizes the known relation
$\tau_q=D_q(q-1)$. We find equation for set of averages related to ordinary,
escort, and generalized probabilities in terms of the deformed expansion as
well. Multifractals related to the Cantor binomial set, exchange currency
series, and porous surface condensates are considered as examples.
 PACS {02.20.Uw, 05.45.Df}
\end{abstract}

\section{Introduction}

Fractal conception \cite{Mandel} has become a widespread idea in contemporary
science (see Refs. \cite{2,Feder,Sor} for review). Characteristic feature of
fractal sets is known to be the self-similarity: if one takes a part of the
whole set, it looks like the original set after appropriate scaling. Formal
basis of the self-similarity is the power-law function $F\sim\ell^{h}$ with the
Hurst exponent $h$ (for time series, value $F$ is reduced to the fluctuation
amplitude and $\ell$ is the interval size within which this amplitude is
determined). While the simple case of monofractal is characterized by a single
exponent $h$, a multifractal system is described by a continuous spectrum of
exponents, singularity spectrum $h(q)$ with argument $q$ being the exponent
deforming measures of elementary boxes that cover the fractal set \cite{multi}.
On the other hand, the parameter $q$ represents a self-similarity degree of a
homogeneous function being intrinsic in self-similar systems \cite{Erzan} (in
this way, within nonextensive thermostatistics, this exponent expresses the
escort probability $P_i\propto p_i^q$ in terms of the original one $p_i$
\cite{T,BS}). In physical applications, a key role is played by the partition
function $Z_q\sim\ell^{\tau(q)}$ with $\ell$ as a characteristic size of boxes
covering multifractal and the exponent $\tau(q)$ connected with the generalized
Hurst exponent $h(q)$ by the relation $\tau(q)=qh(q)-1$.

As fractals are scale invariant sets, it is natural to apply the quantum
calculus to describe multifractals. Indeed, quantum analysis is based on the
Jackson derivative
\begin{equation}
\mathcal{D}_x^\lambda=\frac{\lambda^{x\partial_x}-1}{(\lambda-1)x},\quad
\partial_x\equiv\frac{\partial}{\partial x}
 \label{1}
\end{equation}
that yields variation of a function $f(x)$ with respect to the scaling
deformation $\lambda$ of its argument \cite{QC,Gasper}. First, this idea has
been realized in the work \cite{Erzan} where support space of multifractal has
been proposed to deform by means of action of the Jackson derivative (\ref{1})
on the variable $x$ reduced to the size $\ell$ of covering boxes. In this
letter, we use quite different approach wherein deformation is applied to the
multifractal parameter $q$ itself to vary it by means of finite dilatation
$(\lambda-1)q$ instead of infinitesimal shift ${\rm d}q$. We demonstrate below
that related description allows one to generalize definitions of the partition
function, the mass exponent, and the averages of random variables on the basis
of deformed expansion in power series over difference $q-1$. We apply the
formalism proposed to consideration of multifractals in mathematical physics
(the Cantor binomial set), econophysics (exchange currency series), and solid
state physics (porous surface condensates).

\section{Formulation of the problem}

Following the standard scheme \cite{multi,Feder}, we consider a multifractal
set covered by elementary boxes $i=1,2,\dots,W$ with $W\to\infty$. Its
properties are known to be determined by the partition function
\begin{equation}
Z_q=\sum_{i=1}^W p_i^q
 \label{Z}
\end{equation}
that takes the value $Z_q=1$ at $q=1$, in accordance with the normalization
condition. Since $p_i\leq 1$ for all boxes $i$, the function (\ref{Z})
decreases monotonically from maximum magnitude $Z_q=W$ related to $q=0$ to
extreme values $Z_q\simeq p_{\rm ext}^q$ which are determined in the
$|q|\to\infty$ limit by maximum probability $p_{\rm max}$ on the positive
half-axis $q>0$ and minimum magnitude $p_{\rm min}$ on the negative one. In the
simplest case of the uniform distribution $p_i=1/W$ fixed by the statistical
weight $W\gg 1$, one has the exponential decay $Z_q=W^{1-q}$.

The corner stone of our approach is a generalization of the partition function
(\ref{Z}) by means of introducing a deformation parameter $\lambda$ which
defines, together with the self-similarity degree $q$, {\it a modified
partition function} $\mathcal{Z}_q^\lambda$ reduced to the standard form $Z_q$
at $\lambda=1$. To find the explicit form of the function
$\mathcal{Z}_q^\lambda$ we expand the difference
$\mathcal{Z}_q^\lambda-Z_{\lambda}$ into the deformed series over powers of the
difference $q-1$:
\begin{equation}
\mathcal{Z}_q^\lambda:=Z_{\lambda}
-\sum\limits_{n=1}^\infty\frac{\mathcal{S}_{\lambda}^{(n)}}
{[n]_\lambda!}(q-1)_\lambda^{(n)},\quad Z_{\lambda}=\sum_{i=1}^W p_i^{\lambda}.
 \label{Z1}
\end{equation}
For arbitrary $x$ and $a$, the deformed binomial \cite{QC,Gasper}
\begin{equation} \label{8}
\begin{split}
(x+a)_\lambda^{(n)}&=(x+a)(x+\lambda a)\dots(x+\lambda^{n-1}a)
\\ &=\sum\limits_{m=0}^n\left[{n\atop
m}\right]_\lambda \lambda^{\frac{m(m-1)}{2}}x^m a^{n-m},\ n\geq 1
\end{split}
\end{equation}
is determined by the coefficients $\left[{n\atop
m}\right]_\lambda=\frac{[n]_\lambda!}{[m]_\lambda![n-m]_\lambda!}$ where
generalized factorials $[n]_\lambda!=[1]_\lambda[2]_\lambda\dots[n]_\lambda$
are given by the basic deformed numbers
\begin{equation}
[n]_\lambda=\frac{\lambda^n-1}{\lambda-1}.
 \label{10}
\end{equation}
The coefficients of the expansion (\ref{Z1})
\begin{equation}
\mathcal{S}_{\lambda}^{(n)}=-\left.\big(q\mathcal{D}_q^\lambda\big)^n
Z_q\right|_{q=1},\quad n\geq 1
 \label{Z2}
\end{equation}
are defined by the $n$-fold action of the Jackson derivative (\ref{1}) on the
original partition function (\ref{Z}).

\section{Generalized entropies}

Simple calculations arrive at the explicit expression
\begin{equation}
\mathcal{S}_\lambda^{(n)}=
-\frac{\left[Z_\lambda-1\right]^{(n)}}{(\lambda-1)^n},\quad n\geq 1.
 \label{kernel}
\end{equation}
Hereafter, we use {\it the functional binomial}
\begin{equation}
\left[x_t+a\right]^{(n)}:=\sum\limits_{m=0}^n{n\choose m}x_{t^m}a^{n-m}
 \label{binomial}
\end{equation}
defined with the standard binomial coefficients ${n\choose
m}=\frac{n!}{m!(n-m)!}$ for an arbitrary function $x_t=x(t)$ and a constant
$a$. The definition (\ref{binomial}) is obviously reduced to the Newton
binomial for the trivial function $x_t=t$. The most crucial difference of the
functional binomial from the ordinary one is displayed at $a=-1$ in the limit
$t\to 0$, when all terms of the sum (\ref{binomial}), apart from the first
$x_{t^0}=x_1$, are proportional to $x_{t^m}\to x_0$ to give
\begin{equation}
\lim_{t\to 0}\left[x_t-1\right]^{(n)}=(-1)^n(x_1-x_0).
 \label{limit}
\end{equation}
At $t=1$, one has $\left[x_1-1\right]^{(n)}=0$.

It is easy to see the set of coefficients (\ref{kernel}) is expressed in terms
of the Tsallis entropy \cite{T}
\begin{equation}
S_\lambda=-\sum_i\ln_\lambda(p_i)p_i^\lambda=-\frac{Z_\lambda -1}{\lambda-1}
 \label{S}
\end{equation}
where the generalized logarithm
$\ln_\lambda(x)=\frac{x^{1-\lambda}-1}{1-\lambda}$ is used. As the $\lambda$
deformation grows, this entropy decreases monotonically taking the
Boltzmann-Gibbs form $S_1=-\sum_i p_i\ln(p_i)$ at $\lambda=1$. Obvious equality
\begin{equation}
\mathcal{S}_\lambda^{(n)}=-\frac{\left[(1-\lambda)S_\lambda\right]^{(n)}}
{(\lambda-1)^n},\quad n\geq 1
 \label{K}
\end{equation}
expresses in explicit form the entropy coefficients (\ref{kernel}) in terms of
the Tsallis entropy (\ref{S}) that relates to manifold deformations
$\lambda^m$, $0\leq m\leq n$. At $\lambda=0$ when $Z_0=W$, the limit
(\ref{limit}) gives
$\left[S_0\right]^{(n)}=\left[Z_0-1\right]^{(n)}=(-1)^{n-1}(W-1)$, so that
$\mathcal{S}_{0}^{(n)}=W-1\simeq W$. Respectively, in the limit $\lambda\to 1$
where $S_\lambda\to S_1$ and
$\left[(1-\lambda)S_\lambda\right]^{(n)}\to(1-\lambda)^nS_1^n$, one obtains the
sign-changing values $\mathcal{S}_{1}^{(n)}\to (-1)^{n-1}S_1^n$. Finally, the
limit $|\lambda|\to\infty$ where $S_\lambda\sim\lambda^{-1}$ and
$\left[(1-\lambda)S_\lambda\right]^{(n)}\sim(-1)^n$ is characterized by the
sign-changing power asymptotics
$\mathcal{S}_{\lambda}^{(n)}\sim(-1)^{n-1}\lambda^{-n}$.

For the uniform distribution when $Z_\lambda=W^{1-\lambda}$, the dependence
(\ref{S}) is characterized by the value $S_0\simeq W$ in the limit $\lambda\ll
1$ and the asymptotics $S_\lambda\sim 1/\lambda$ at $\lambda\gg 1$ (in the
point $\lambda=1$, one obtains the Boltzmann entropy $S_1=\ln(W)$). As a
result, with the $\lambda$ growth along the positive half-axis, the
coefficients (\ref{kernel}) decrease from the magnitude
$\mathcal{S}_{0}^{(n)}\simeq W$ to the sign-changing values
$\mathcal{S}_{1}^{(n)}=(-1)^{n-1}\left[\ln(W)\right]^n$ and then tend to the
asymptotics $\mathcal{S}_{\lambda}^{(n)}\sim(-1)^{n-1}\lambda^{-n}$; with the
$|\lambda|$ growth along the negative half-axis, the coefficients
(\ref{kernel}) vary non-monotonic tending to
$\mathcal{S}_{\lambda}^{(n)}\sim-|\lambda|^{-n}$ at $\lambda\to-\infty$.

\section{Generalized fractal dimensions}

Within pseudo-thermodynamic picture of multifractal sets \cite{BS}, effective
values of the free energy $\tau_q$, the internal energy $\alpha$, and the
entropy $f$ are defined as follows:
\begin{equation}\label{fa}
\tau_q=\frac{\ln(Z_q)}{\ln(\ell)},\ \alpha=\frac{\sum_i P_i\ln
p_i}{\ln(\ell)},\ f=\frac{\sum_i P_i\ln P_i}{\ln(\ell)}.
\end{equation}
Here, $\ell\ll 1$ stands for a scale, $p_i$ and $P_i$ are original and escort
probabilities connected with the definition
\begin{equation}\label{prob}
P_i(q)=\frac{p_i^q}{\sum_i p_i^q}=\frac{p_i^q}{Z_q}.
\end{equation}
Inserting the last equation into the second expression (\ref{fa}), one obtains
the Legendre transform $\tau_q=q\alpha_q-f(\alpha_q)$ where $q$ plays the role
of the inverse temperature and the internal energy is specified with the state
equation $\alpha_q=\frac{{\rm d}\tau_q}{{\rm d}q}$ \cite{Feder}. It is easy to
convince the escort probability (\ref{prob}) is generated by the mass exponent
given by the first definition (\ref{fa}) with accounting eq. (\ref{Z}):
\begin{equation}\label{P}
qP_i(q)=\ln(\ell)p_i\frac{\partial\tau_q}{\partial
p_i}=\frac{\partial\ln(Z_q)}{\partial \ln(p_i)}.
\end{equation}

Along the line of the generalization proposed, we introduce further {\it a
deformed mass exponent} $\tau_q^\lambda$ related to the original one $\tau_q$
according to the condition $\tau_q=\lim_{\lambda\to 1}\tau_q^\lambda$. By
analogy with eq. (\ref{Z1}), we expand this function into the deformed series
\begin{equation}
\tau_q^\lambda:=
\sum\limits_{n=1}^\infty\frac{D_{\lambda}^{(n)}}{[n]_\lambda!}(q-1)_\lambda^{(n)}
 \label{tau}
\end{equation}
being the generalization of the known relation $\tau_q=D_q(q-1)$ connecting the
mass exponent $\tau_q$ with the multifractal dimension spectrum $D_q$
\cite{Feder}. Similarly to eqs. (\ref{Z2}, \ref{kernel}), the coefficients
$D_{\lambda}^{(n)}$ are expressed in the form
\begin{equation}
D_\lambda^{(n)}=\left.\big(q\mathcal{D}_q^\lambda\big)^n
\tau_q\right|_{q=1}=\frac{\left[\tau_\lambda-1\right]^{(n)}}{(\lambda-1)^n},\quad
n\geq 1
 \label{DD}
\end{equation}
where the use of the definition (\ref{binomial}) assumes the term with $m=0$
should be suppressed because $\tau_{\lambda^0}=0$. At $n=1$, the last equation
(\ref{DD}) is obviously reduced to the ordinary form
$D_\lambda^{(1)}=\tau_{\lambda}/(\lambda-1)$, while the coefficients
$D_{\lambda}^{(n)}$ with $n>1$ include terms proportional to $\tau_{\lambda^m}$
to be related to manifold deformations $\lambda^m$, $1<m\leq n$. To this end,
the definition (\ref{DD}) yields a hierarchy of the multifractal dimension
spectra related to multiplying deformations of different powers $n$.

Making use of the limit (\ref{limit}), where the role of a function $x_t$ is
played by the mass exponent $\tau_\lambda$ with $\tau_0=-1$ and $\tau_1=0$,
gives the value $\left[\tau_0-1\right]^{(n)}=(-1)^n$ at the point $\lambda=0$,
where the coefficients (\ref{DD}) take the magnitude $D_0^{(n)}=1$ related to
the dimension of the support segment. In the limits $\lambda\to\pm\infty$,
behavior of the mass exponent $\tau_\lambda\simeq D_{\rm ext}\lambda$ is
determined by extreme values $D_{\rm ext}$ of the multifractal dimensions which
are reduced to the minimum magnitude $D_{\rm min}=D_{\infty}^{(n)}$ and the
maximum one $D_{\rm max}=D_{-\infty}^{(n)}$ \cite{Feder}. On the other hand, in
the limits $\lambda\to\pm\infty$, the extreme values $Z_{\lambda}\simeq p_{\rm
ext}^\lambda$ of the partition function (\ref{Z}) are determined by related
probabilities $p_{\rm ext}$. As a result, the first definition (\ref{fa}) gives
the mass exponents $\tau_\lambda\simeq\lambda\frac{\ln(p_{\rm
ext})}{\ln(\ell)}$ and the coefficients (\ref{DD}) tend to the minimum
magnitude $D_\infty^{(n)}\simeq\frac{\ln(p_{\rm max})}{\ln(\ell)}$ at
$\lambda\to\infty$ and the maximum one $D_{-\infty}^{(n)}\simeq\frac{\ln(p_{\rm
min})}{\ln(\ell)}$ at $\lambda\to-\infty$.

For the uniform distribution whose partition function is
$Z_\lambda=W^{1-\lambda}$, the expression $Z_\lambda:=\ell^{\tau_\lambda}$
gives the fractal dimension $D=\frac{\ln(W)}{\ln(1/\ell)}$ which tends to $D=1$
when the size of covering boxes $\ell$ goes to the inverse statistical weight
$1/W$. Being unique, this dimension relates to a monofractal with the mass
exponent $\tau_\lambda=D(\lambda-1)$ whose insertion into the definition
(\ref{DD}) arrives at the equal coefficients $D_\lambda^{(n)}=D$ for all orders
$n\geq 1$.

\section{Relations between generalized entropies and fractal dimensions}

Since either of the deformed series (\ref{Z1}) and (\ref{tau}) describes a
multifractal completely, their coefficients should be connected in some way. It
is easy to find an explicit relation between the first of these coefficients
$S_\lambda=\ln_\lambda\left(Z_\lambda^{\frac{1}{1-\lambda}}\right)$ and
$D_\lambda=\frac{\ln(Z_\lambda)}{(\lambda-1)\ln(\ell)}$ being the Tsallis
entropy $S_{\lambda}=\mathcal{S}_{\lambda}^{(1)}$ and the multifractal
dimension $D_{\lambda}=D_{\lambda}^{(1)}$. The use of the relation
$Z_\lambda=\ell^{\tau_\lambda}$, the connection
$\tau_\lambda=D_\lambda(\lambda-1)$, and the Tsallis exponential
$\exp_\lambda(x)=\left[1+(1-\lambda)x\right]^{\frac{1}{1-\lambda}}$ arrives at
the expressions
\begin{equation}
S_\lambda=\ln_\lambda\left(W^{D_\lambda}\right),\quad
D_\lambda=\frac{\ln\left[\exp_\lambda\left(S_\lambda\right)\right]}{\ln(W)}
 \label{SDSD}
\end{equation}
where the statistical weight $W=1/\ell$ is used. Unfortunately, it is
impossible to set any closed relation between the coefficients (\ref{kernel})
and (\ref{DD}) at $n>1$. However, the use of the partition function
$Z_\lambda=W^{-D_\lambda(\lambda-1)}$ allows us to write
\begin{equation}
D_\lambda^{(n)}=\frac{\left[D_\lambda(\lambda-1)-1\right]^{(n)}}{(\lambda-1)^n},
 \label{lambda}
\end{equation}
\begin{equation}
\mathcal{S}_{\lambda}^{(n)}=-\frac{\left[W^{-D_\lambda(\lambda-1)}-1\right]^{(n)}}
{(\lambda-1)^n}.
 \label{entropy}
\end{equation}
Thus, knowing the first coefficients of expansions (\ref{Z1}) and (\ref{tau})
connected with the relations (\ref{SDSD}), one can obtain them for arbitrary
orders $n>1$.

\section{Random variable  distributed over a multifracal set}

Let us consider an observable $\phi_i$ distributed over a multifracal set with
the average $\left<\phi\right>_q^\lambda=\sum_i
\phi_i\mathcal{P}_i(q,\lambda)$. Related probability is determined by the
equation
\begin{equation} \label{PP}
\lambda\mathcal{P}_i(q,\lambda):=p_i+\ln(\ell)p_i\frac{\partial\tau_q^\lambda}{\partial
p_i}
\end{equation}
that generalizes eq. (\ref{P}) for the escort probability due to the $\lambda$
deformation. With accounting eqs. (\ref{P} -- \ref{DD}), this average can be
expressed in terms of the deformed series
\begin{equation} \label{O}
\lambda\left<\phi\right>_q^\lambda=\left<\phi\right>+\sum\limits_{n=1}^\infty
\frac{\left[\lambda\left<\phi\right>_\lambda-1\right]^{(n)}}{[n]_\lambda!(\lambda-1)^{n}}
(q-1)_\lambda^{(n)}
\end{equation}
where $\left<\phi\right>=\sum_i\phi_i p_i$ and
$\left<\phi\right>_\lambda=\sum_i\phi_i P_i(\lambda)$. This equation allows one
to find the mean value $\left<\phi\right>_q^\lambda$ versus the self-similarity
degree $q$ at the $\lambda$ deformation fixed. In the case $\lambda=q$ when
definition (\ref{8}) gives $(q-1)_q^{(n)}=0$ for $n>1$, the equation (\ref{O})
arrives at the connection $\left<\phi\right>_q^q=\left<\phi\right>_q$ between
mean values related to generalized and escort probabilities given by eqs.
(\ref{PP}) and (\ref{P}), respectively. At the point $\lambda=0$ where
according to eqs. (\ref{8}, \ref{10}, \ref{limit})
$(q-1)_0^{(n)}=(q-1)q^{n-1}$, $[n]_0!=1$ and
$\left[\lambda\left<\phi\right>_\lambda-1\right]^{(n)}\to(-1)^n\left<\phi\right>$,
eq. (\ref{O}) is reduced to identity; here, one has the uniform distribution
$P_i(q,0)=1/W$ for an arbitrary $p_i$ and the average is
$\left<\phi\right>_q^0=W^{-1}\sum_i \phi_i$. At $\lambda=1$ when
$\left[\left<\phi\right>_1-1\right]^{(n)}=0$, the equation (\ref{O}) arrives at
the ordinary average $\left<\phi\right>=\sum_i\phi_i p_i$ because the
distribution $P_i(q,1)$ is reduced to $p_i$. Setting $\sum_{n=1}^\infty
(-1)^{n-1}\left<\phi\right>_{\lambda^n}\to\left<\phi\right>_\infty$ for
$\lambda\gg 1$ where
$(q-1)_\lambda^{(n)}\sim(-1)^{n-1}(q-1)\lambda^{n(n-1)/2}$,
$[n]_\lambda!\sim\lambda^{n(n-1)/2}$ and
$\left[\lambda\left<\phi\right>_\lambda-1\right]^{(n)}\sim\lambda^n\left<\phi\right>_{\lambda^n}$,
one obtains the simple dependence
$\lambda\left<\phi\right>_q^\lambda=\left<\phi\right>+(q-1)\left<\phi\right>_\infty$,
according to which the average $\left<\phi\right>_q=\left<\phi\right>_q^q$
tends to the limit $\left<\phi\right>_\infty$ with the $q$-growth.

\section{Examples}

\begin{figure}[!h]
\centering
 \includegraphics[width=70mm]{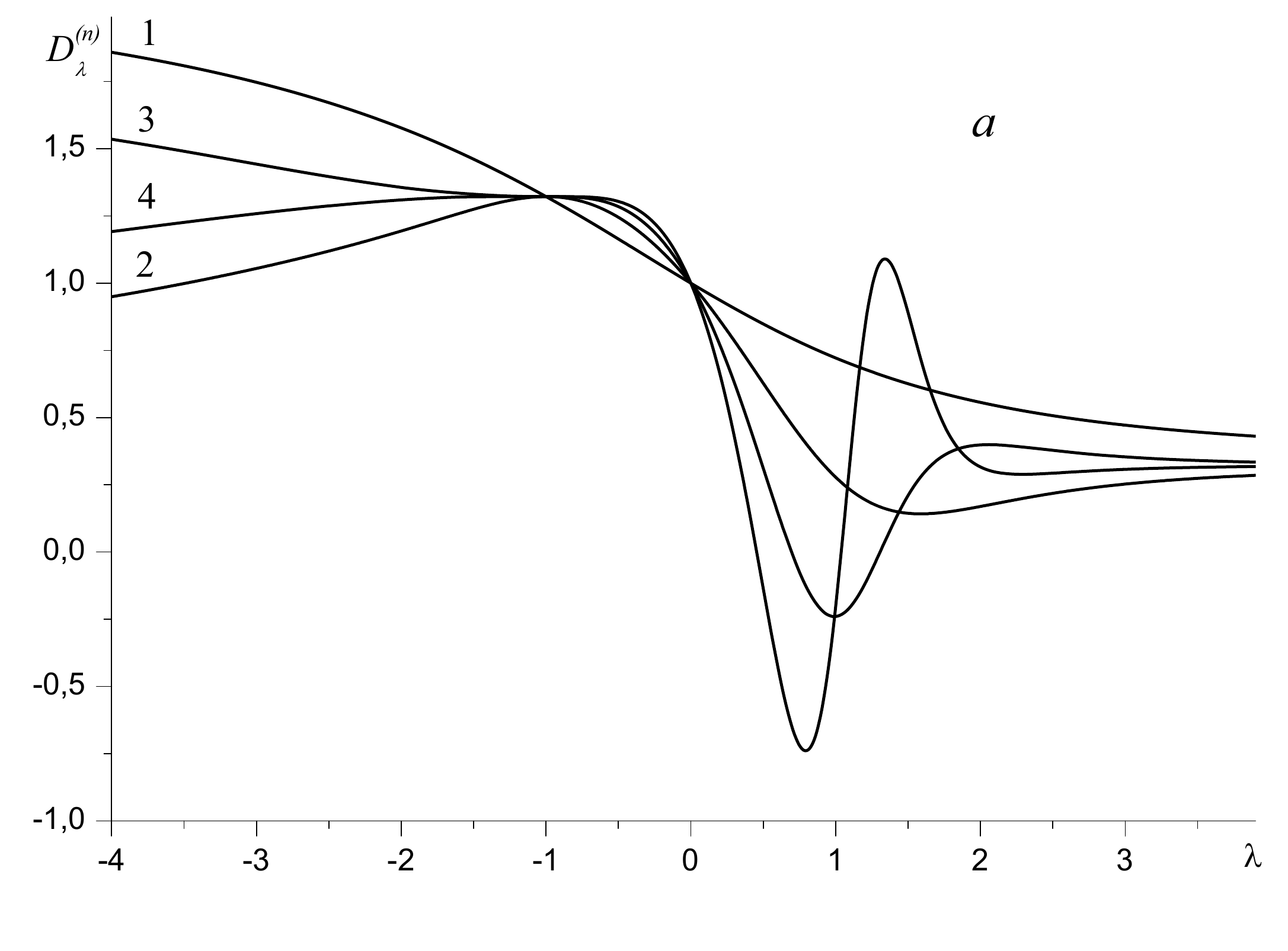}\\
 \includegraphics[width=70mm]{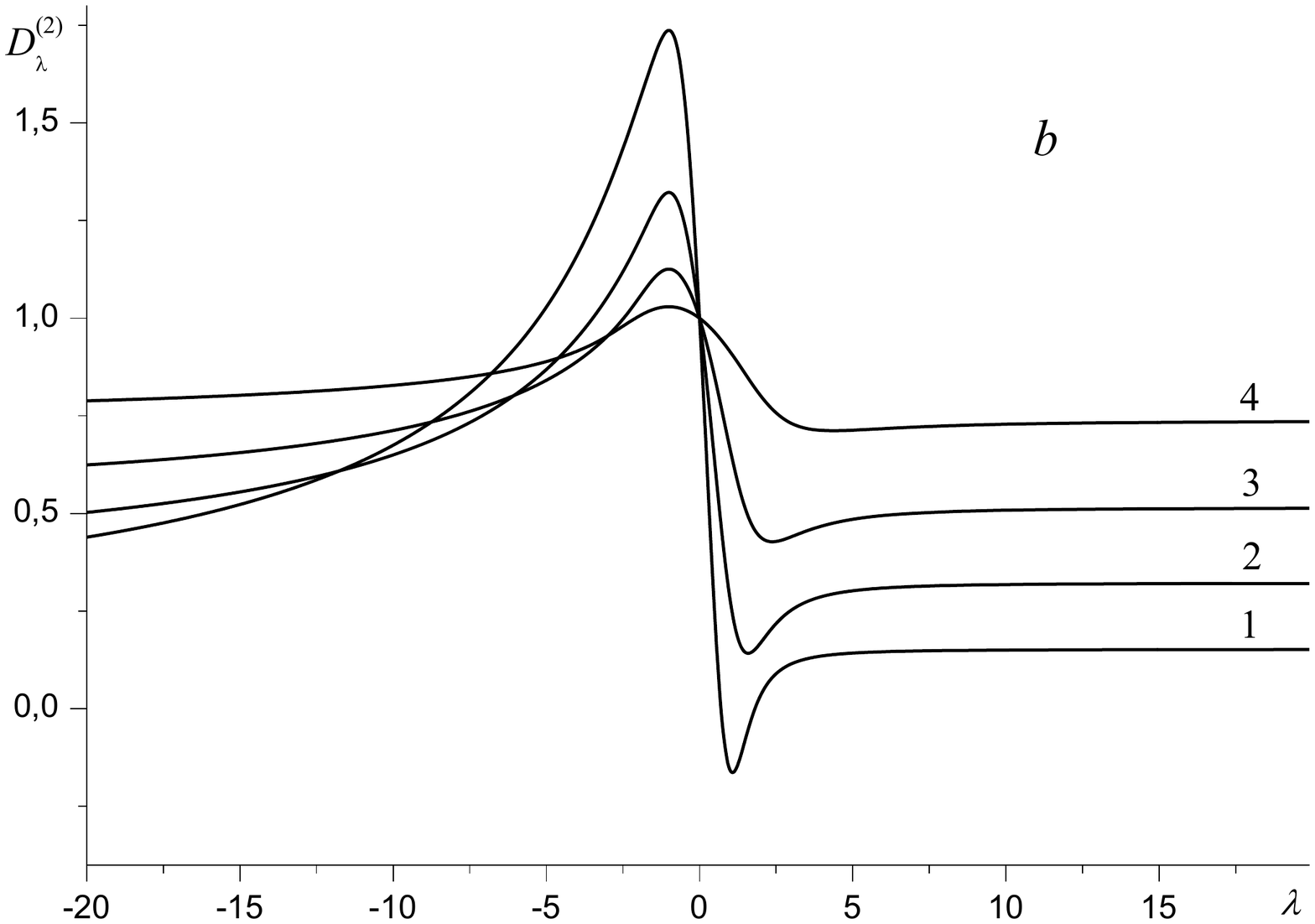}
  \caption{Fractal dimension coefficients (\ref{lambda}) versus deformation
  of the Cantor binomial set at $p=0.2$ (a) and $n=2$ (b)
  (curves 1 -- 4 correspond to $n=1, 2, 3, 4$ on the upper panel and $p=0.1, 0.2, 0.3, 0.4$
  on the lower one).}
 \label{Dimensions}
\end{figure}

\begin{figure}[!h]
 \centering
 \includegraphics[width=70mm]{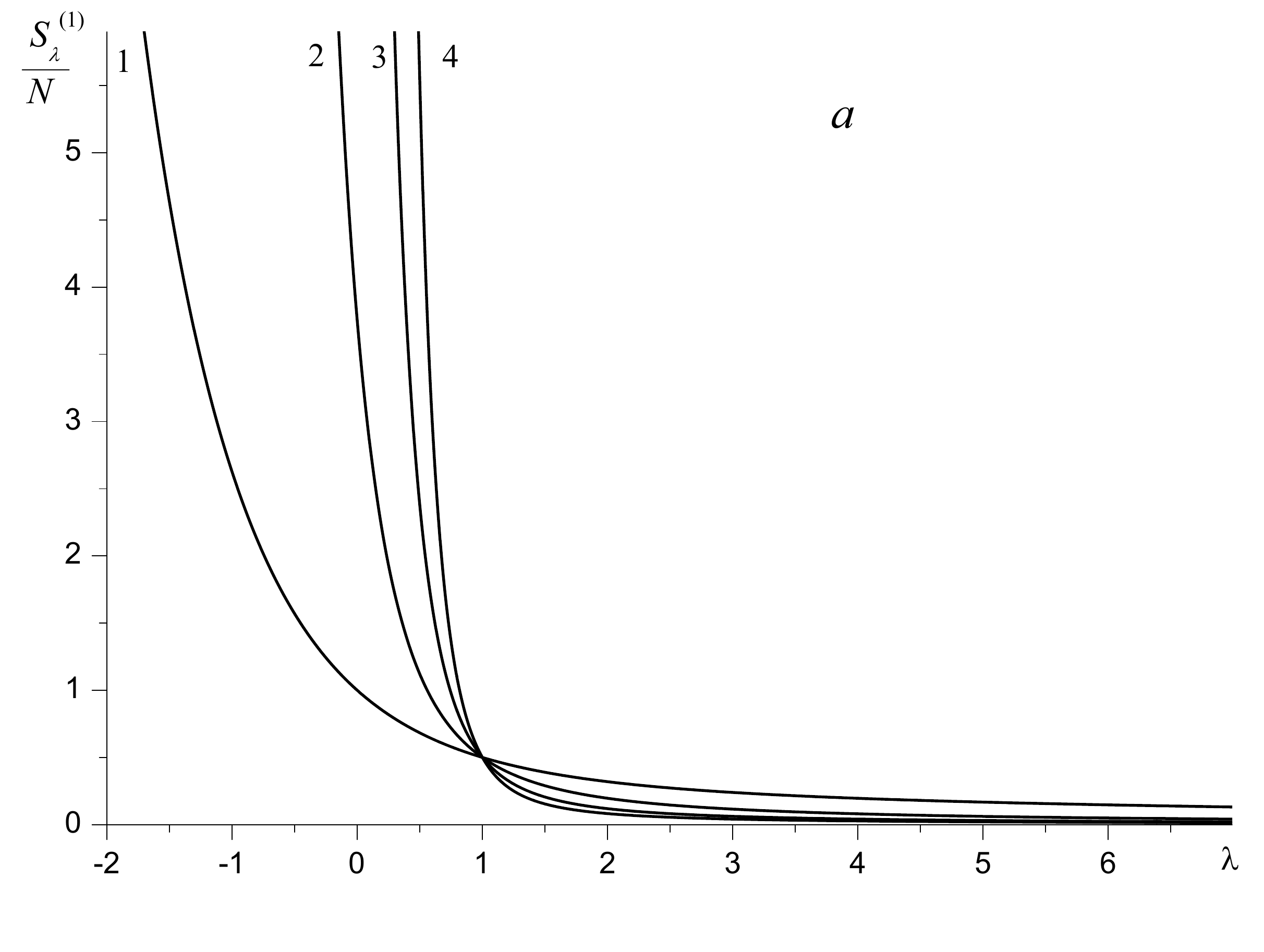}\\
 \includegraphics[width=70mm]{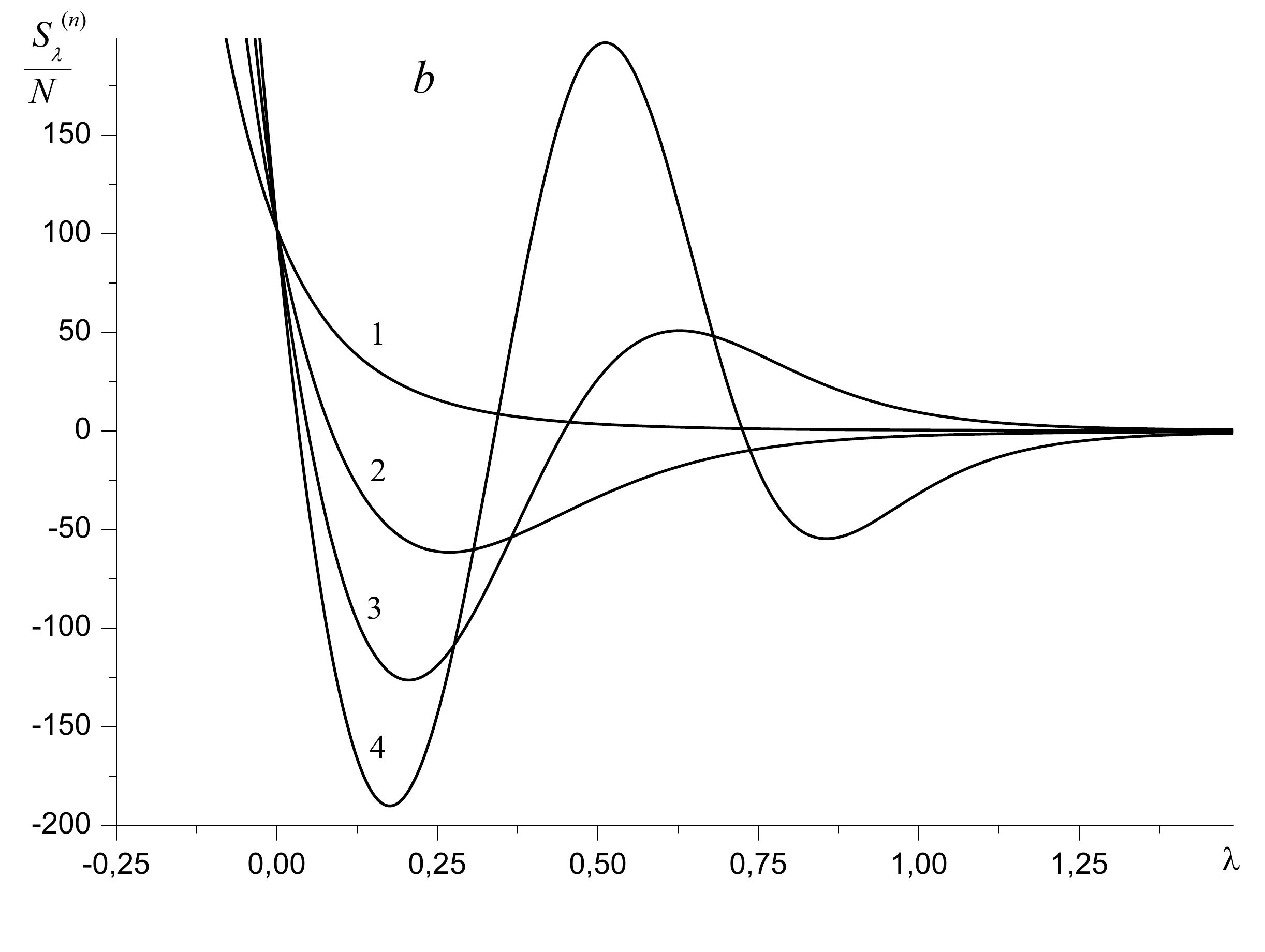}
  \caption{Effective entropies (\ref{entropy})
  of the Cantor binomial set at $p=0.2$ and:
  a) $n=1$ (curves 1 -- 4 correspond to $N=1,4,8,12$);
  b) $N=10$ (curves 1 -- 4 correspond to $n=1, 2, 3, 4$).}
 \label{Entropies}
\end{figure}

To demonstrate the approach developed we consider initially the simplest
example of the Cantor binomial set \cite{Feder}. It is generated by the
$N$-fold dividing of the unit segment into equal parts with elementary lengths
$\ell=(1/2)^N$, then each of these is associated with binomially distributed
products $p^m(1-p)^{N-m}$, $m=0,1,\dots,N$ of probabilities $p$ and $1-p$. In
such a case, the partition function (\ref{Z}) takes the form
$Z_q=\left[p^{q}+(1-p)^{q}\right]^N$ to be equal $Z_q=\ell^{\tau_q}$ with the
mass exponent $\tau_q=\frac{\ln\left[p^{q}+(1-p)^{q}\right]}{\ln(1/2)}$
\cite{Feder}. Related dependencies of the fractal dimension coefficients
(\ref{lambda}) on the deformation parameter are depicted in Figure
\ref{Dimensions} for different orders $n$ and probabilities $p$.
As the upper panel shows, at the $p$ given the monotonic decay of the fractal
dimension $D_\lambda^{(n)}$, being usual at $n=1$, transforms into the
non-monotonic dependencies, whose oscillations are the stronger the more the
order $n$ is. According to the lower panel, such behavior is kept with
variation of the $p$ probability, whose growth narrows the dimension spectrum.
In contrast to the fractal dimensions (\ref{lambda}), the entropy coefficients
(\ref{entropy}) depend on the effective number of particles $N$. This
dependence is demonstrated by example of the Tsallis entropy
$S_{\lambda}=\mathcal{S}_{\lambda}^{(1)}$ depicted in Figure \ref{Entropies}a:
with the deformation growth, this entropy decays
the faster the more $N$ is (by this, the specific value $S_{\lambda}/N$ remains
constant at $\lambda=1$). According to Figure \ref{Entropies}b, with increase
of the order $n$, the monotonically decaying dependence
$\mathcal{S}_{\lambda}^{(1)}$ transforms into the non-monotonic one,
$\mathcal{S}_{\lambda}^{(n)}$, whose oscillations grow with $n$.
Characteristically, for arbitrary values $n$ and $p$, the magnitude
$\mathcal{S}_0^{(n)}$ (being equal $2^N$ for the Cantor binomial set) remains
constant.

\begin{figure}[!h]
 \centering
 \includegraphics[width=70mm]{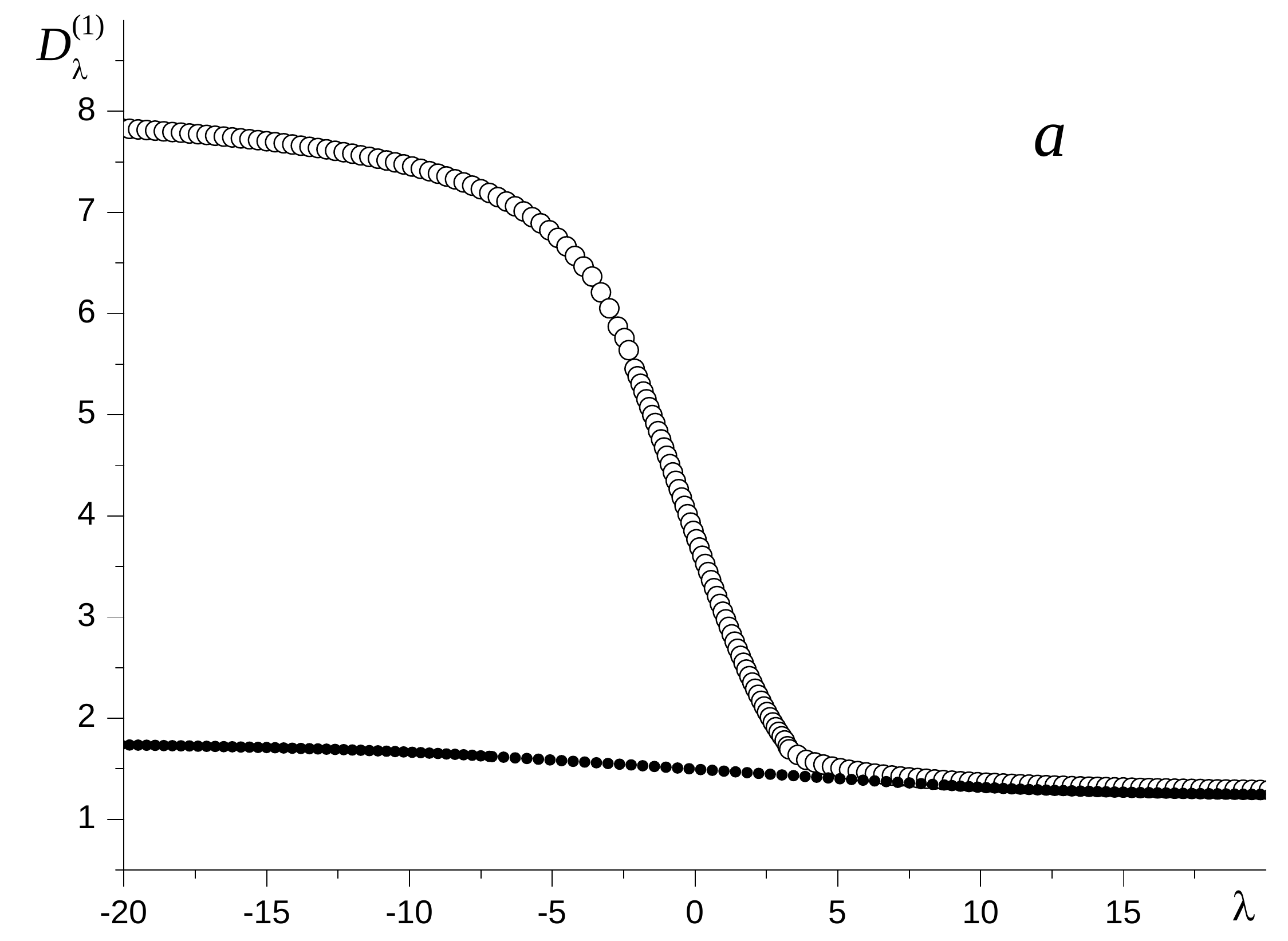}\\
 \includegraphics[width=70mm]{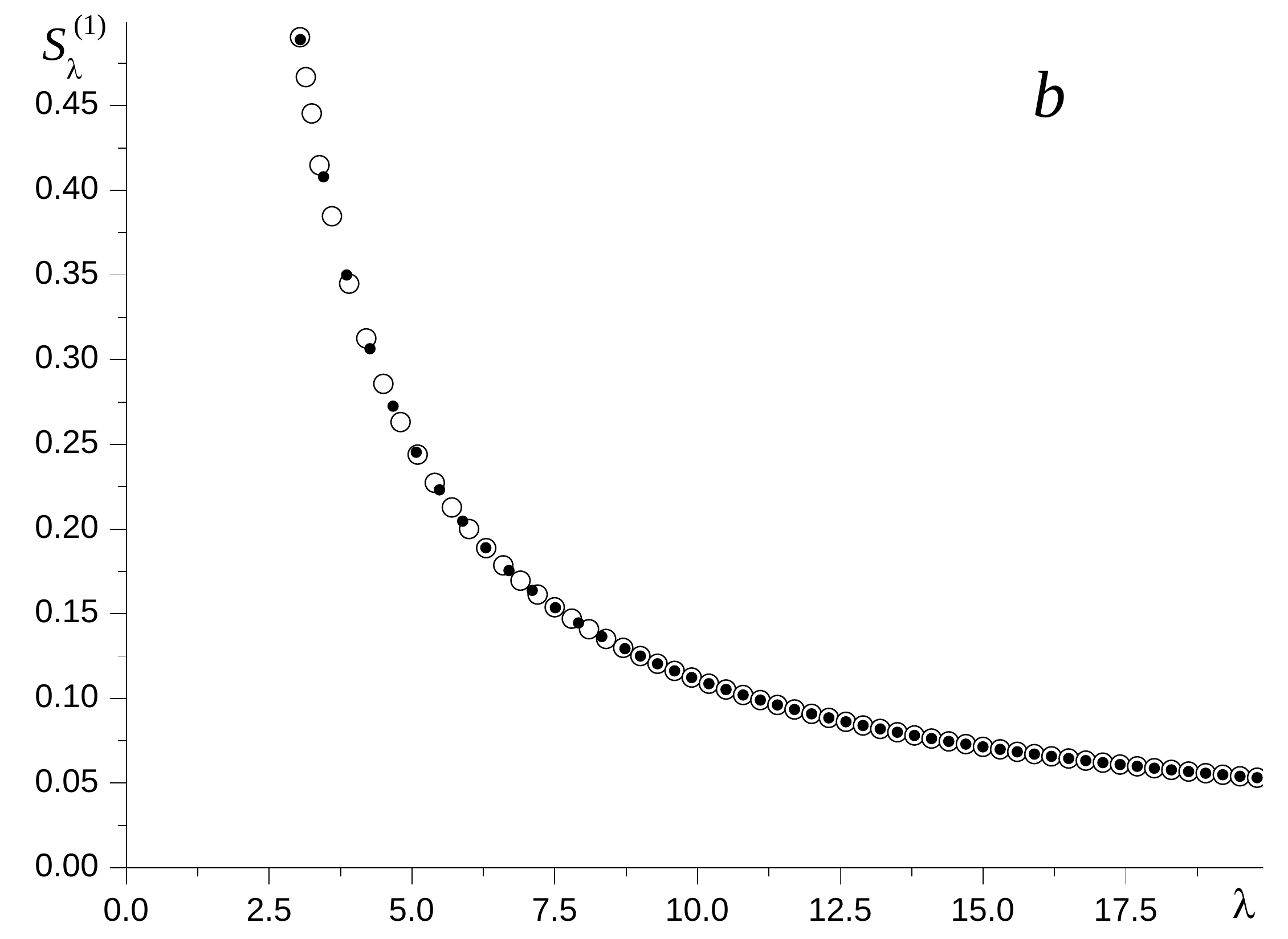}\\
\includegraphics[width=70mm]{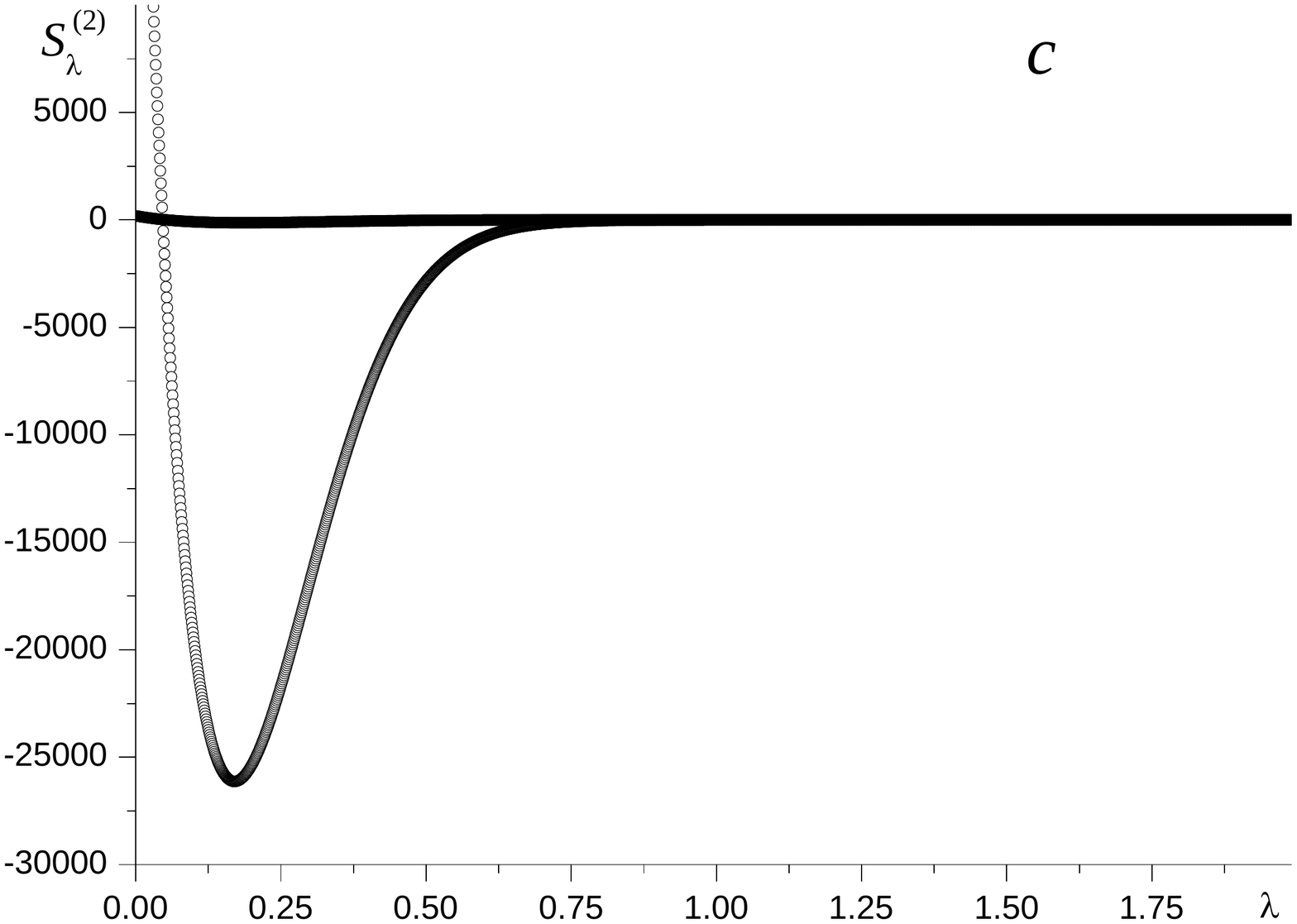}
 \caption{a) Fractal dimension coefficients (\ref{lambda})
 at $n=1$ for the time series of the currency exchange of
 euro to US dollar; related entropy coefficients (\ref{entropy})
 at $n=1$ (b) and  $n=2$ (c) (bold dots correspond to the
 time interval before financial crisis, circles -- after crisis).}
 \label{currency}
\end{figure}

As the second example we consider the time series of the currency exchange of
euro to US dollar in the course of the 2007 -- 2009 years which include
financial crisis (data are taken from website www.fxeuroclub.ru). To ascertain
the crisis effect we study the time series intervals before (January, 2007 --
May, 2008) and after crisis (June, 2008 -- October, 2009).\footnote{The point
of the crisis is fixed under condition of maximal value of the time series~dispersion.}
 Moreover, we restrict ourselves by consideration of the
coefficients (\ref{lambda}) and (\ref{entropy}) of the lowest orders $n$ which
make it possible to visualize a difference between fractal characteristics of
the time series intervals pointed out. Along this way, we base on the method of
the multifractal detrended fluctuation analysis \cite{Kantel} to find the mass
exponent $\tau(q)$, whose use arrives at the dependencies depicted in Figure
\ref{currency}.
Comparison of the data taken before and after financial crisis shows that it
affects already on the fractal dimension coefficient of the lowest order $n=1$,
but has not an effect on the Tsallis entropy related to $n=1$, while the
entropy coefficient of the second order $n=2$ is found to be strongly sensible
to the crisis. Thus, this example demonstrates visually that generalized
multifractal characteristics elaborated on the basis of the developed formalism
allow one to study subtle details of self-similarly evolved processes.

\begin{figure}[!h]
 \centering

 a\hspace{5.5cm}b\\
 \vspace{0.2cm}
 \includegraphics[width=120mm]{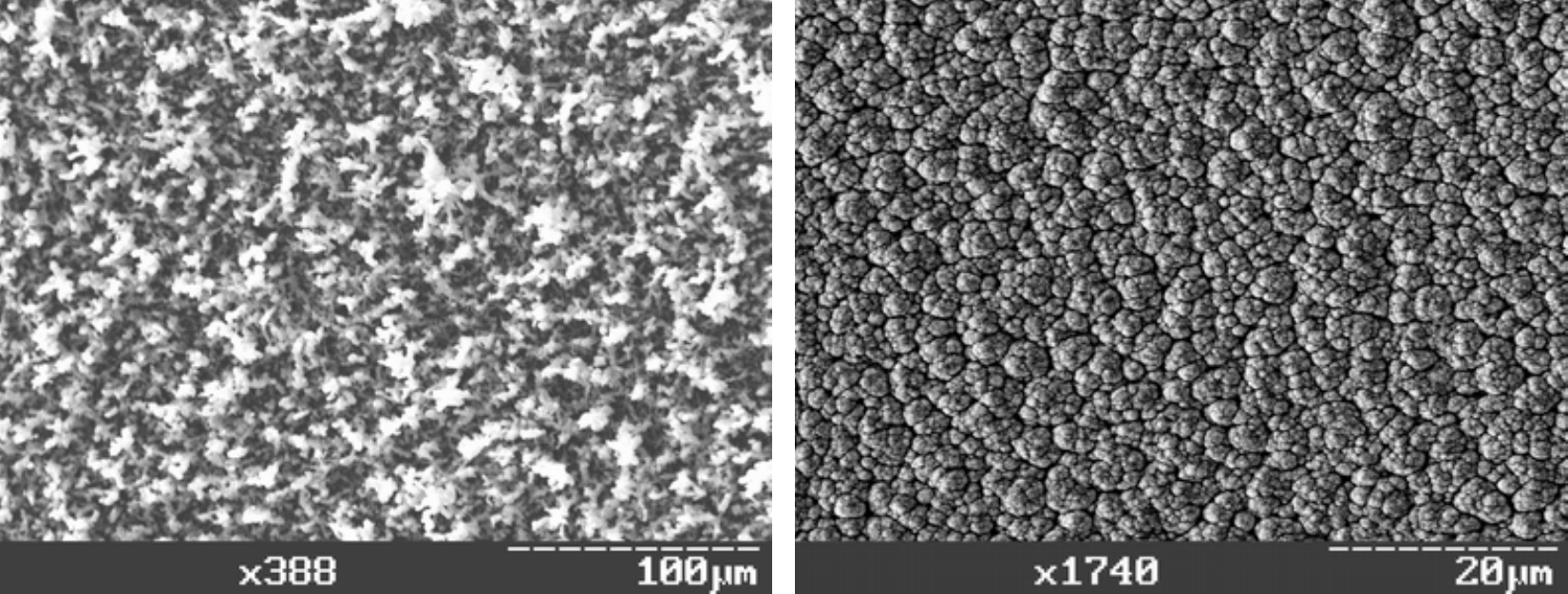}\\
 \vspace{0.2cm}
  c\hspace{5.5cm}d\\
  \includegraphics[width=60mm]{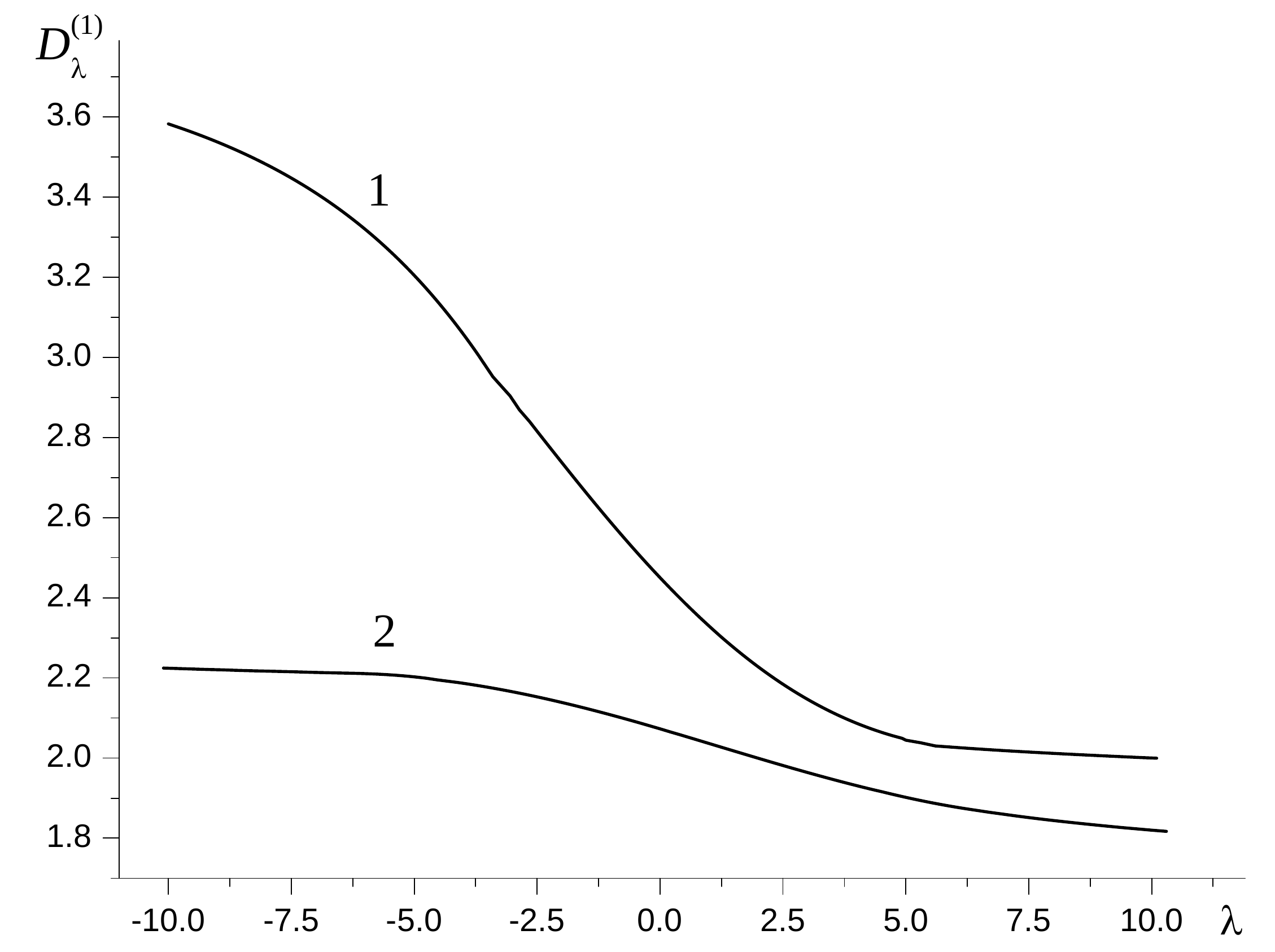}\includegraphics[width=60mm]{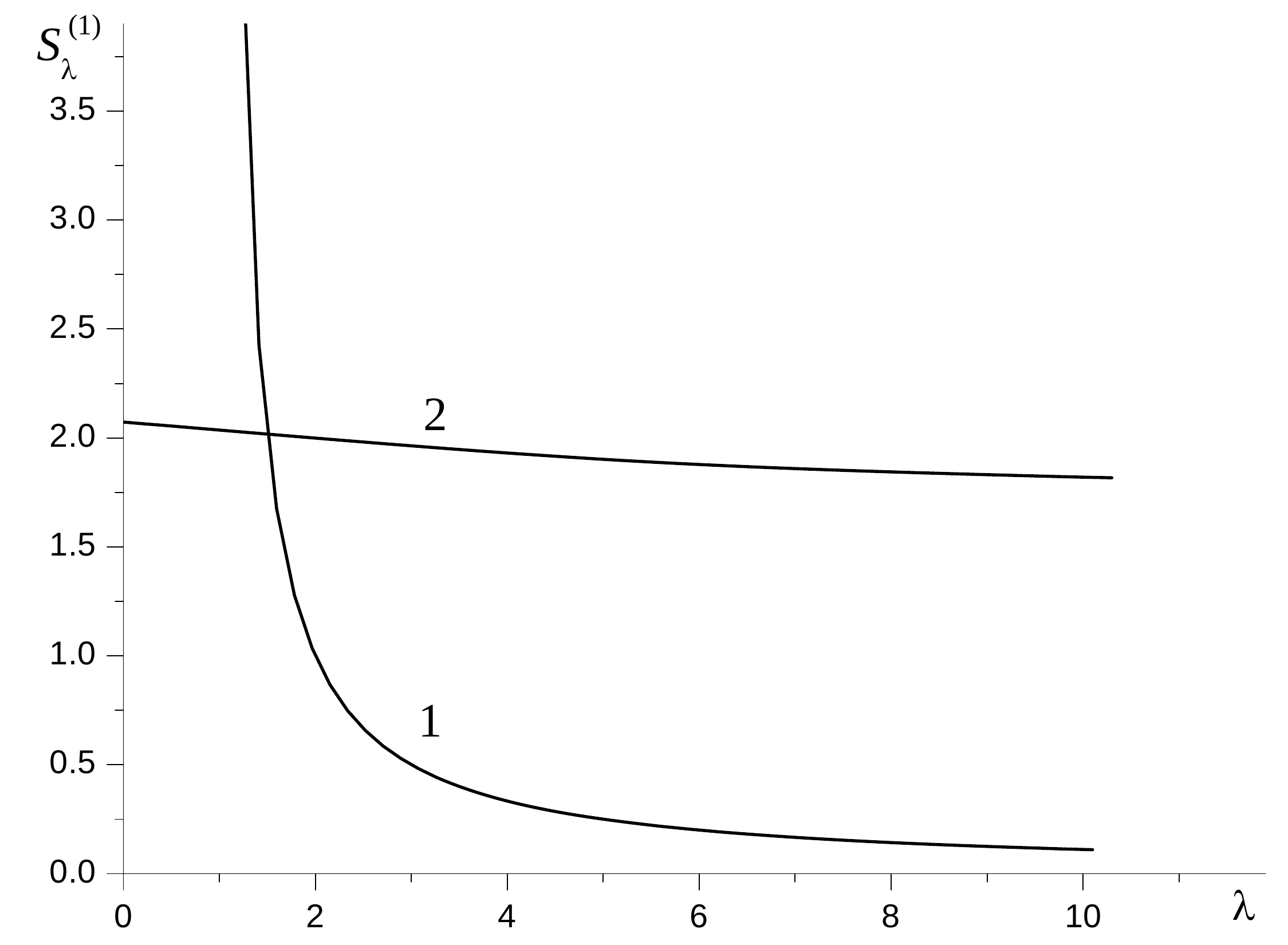}
  \caption{Scanning electron microscopy images of {\it ex-situ} grown carbon (a)
  and titanium (b) condensates; corresponding fractal dimensions (c) and entropies (d)
  at $n=1$ (curves 1, 2 relate to  carbon and titanium).}
 \label{surface}
\end{figure}

The last example is concerned with macrostructure of condensates which have
been obtained as result of sputtering of substances in accumulative ion-plasma
devices \cite{Perekr}. The peculiarity of such a process is that its use has
allowed to obtain the porous condensates type of those that are shown in
Figures \ref{surface}a and \ref{surface}b for carbon and titanium,
respectively.
These condensates are seen to have apparent fractal macrostructure, whose
handling gives the fractal dimension spectra and the entropies depicted in
Figures \ref{surface}c and \ref{surface}d, respectively. Comparing these
dependencies, we can convince that difference between the carbon condensate,
which has strongly rugged surface, with the titanium one, becomes apparent
already at use of the fractal dimension and entropy coefficients
(\ref{lambda}), (\ref{entropy}) related to the lowest order $n=1$. Thus, in the
case of multifractal objects with strongly different structures the use of
usual multifractal characteristics is appeared to be sufficient.

\section{Conclusion}

Generalizing the multifractal theory \cite{multi}, we have represented the
partition function (\ref{Z1}), the mass exponent (\ref{tau}), and the average
$\left<\phi\right>_q^\lambda$ of self-similarly distributed random variable as
deformed series in power of difference $q-1$. Coefficients of these expansions
are shown to be determined by the functional binomial (\ref{binomial}) that
expresses both multifractal dimension spectra (\ref{lambda}) and generalized
Tsallis entropies (\ref{entropy}) related to manifold deformations. We have
found eq. (\ref{O}) for the average related to the generalized probability
(\ref{PP}) subject to the deformation. Recently, making use of above formalism
has allowed us to develop a field theory for self-similar statistical systems
\cite{OSh}.

As examples of multifractal sets in the mathematical physics, objects of the
solid state physics, and processes in the econophysics, we have applied the
formalism developed to consideration of the Cantor manifolds, porous surface
condensates, and exchange currency series, respectively. The study of the
Cantor set has shown that both fractal dimension coefficients (\ref{lambda})
and entropies (\ref{entropy}) coincide with usual multifractal characteristics
in the lowest order $n=1$, but display very complicated behaviour at $n>1$. On
the contrary, consideration of both carbon and titanium surface condensates has
shown that their macrostructures can be characterized by use of usual fractal
dimension and entropy coefficients related to the order $n=1$. Much more
complicated situation takes place in the case of the time series type of
currency exchange series. Here, a difference between various series is
displayed for the fractal dimension coefficients already in the lowest order
$n=1$, but the entropy coefficients coincide at $n=1$, but become different at
$n=2$. This example demonstrates the need to use generalized multifractal
characteristics obtained within framework of the formalism developed.

\end{document}